\documentstyle[12pt,psfig]{article}

\begin{document} 

\title{$\rho$-Meson Production and Decay in Proton-Nucleus 
Collisions\thanks{Supported by Forschungszentrum J\"ulich}}
\author{A. Sibirtsev and W. Cassing \\
Institut f\"ur Theoretische Physik, Universit\"at Giessen \\
D-35392 Giessen, Germany}
\maketitle
\date{ }
\vspace{2cm}
\vspace{1cm}

\begin{abstract}
We analyze the production of $\rho$-mesons in $p + A$ reactions including
both the production by proton-nucleon as well as pion-nucleon collisions
within a coupled channel transport approach. The final state interactions
of the $\rho$-meson with nucleons are evaluated from a resonance model
which allows to extract elastic and inelastic cross sections. We include
the latter final state interactions, the $\rho$-meson decay into two pions
as well as the final pion-nucleon interactions within the transport
approach. We find the invariant mass distribution of pion pairs to be
sensitive to the $\rho$-meson properties in the nuclear medium. 
However, due
to the strong final state interactions of pions only light targets like
$^{12}C$ might be suited to extract  a $\rho$-signal from the uncorrelated
two pion background which carries information about the
in-medium properties of the $\rho$-meson.
\end{abstract}

\newpage
 
\section{Introduction}
The restoration of chiral symmetry at high baryon density or temperature
nowadays is a question of primary interest. As 
suggested a long time ago the
short lived $\rho$-meson is a primary candidate 
to study possible in-medium
mass shifts or dispersion relations as e.g. predicted by 
Brown - Rho scaling \cite{Brown} or  QCD sum-rules \cite{Hatsuda}. 
Whereas the dileptonic
decay of the $\rho$-meson appears as an ideal 
probe of the in-medium properties
of the $\rho$-meson, the two pion hadronic decay in principle carries the
same information provided that the decay products suffer only little
rescattering. 

At SPS energies the dilepton data of the CERES \cite{CERES} and HELIOS-3 
Collaborations \cite{HELIOS} have indicated a possible mass shift of the
$\rho$-meson following the analysis of Li, Ko and Brown \cite{Li} as well
as Cassing et al. \cite{Ca95,Ca96,Ca97}. A recent summary of the situation
has been given by Drees \cite{Drees}. However, the approaches in \cite{Li}
and \cite{Ca95} differ in the actual assumptions on the $\rho$-meson mass 
shift. In the model of Ref. \cite{Li} a dropping of the $\rho$-mass at
normal nuclear matter density ${\rho}_0$ of about 
30-35\% is employed whereas in the
transport study of Ref. \cite{Ca95,Ca96,Ca97} the $\rho$-mass drops only
by about 18\%  at ${\rho}_0$ in line with the QCD sum-rule study of 
Hatsuda and Lee \cite{Hatsuda}. Such drastic in-medium effects 
should also be seen in
proton-nucleus or pion-nucleus collisions \cite{Schoen,Ca97a} via the
electromagnetic decay of the $\rho$-meson or even the $\omega$-meson 
in case
of $\pi$ + A reactions \cite{Schoen}.

In the present paper we explore the possibility if such medium effects of
the $\rho$-meson might also be measured by two-pion invariant mass 
distributions in proton-nucleus reactions. Our study thus is 
organized as follows:
In Section 2 we present the elementary $\rho$-meson production 
cross sections from nucleon-nucleon and pion-nucleon collisions 
and determine the elastic
and inelastic cross sections of the $\rho$-mesons 
with nucleons from a resonance
model. In Section 3 we discuss possible in-medium modifications of the
$\rho$-meson as incorporated in the present simulations.
In Section 4 we test our transport approach with respect to pion spectra
in proton-nucleus reactions. Section 5 is devoted to the actual $\rho$
and pion dynamics in $p+{^{12}C}$ reactions at 2 - 2.5 GeV as well as to
the reconstruction of the $\rho$-mass distribution from two-pion invariant
mass spectra, while Section 6 concludes our work with a summary 
and discussion of open problems. 

\section{Elementary cross sections}
\subsection{$\rho$-meson production}
In a previous study we have investigated the production
of vector mesons from proton-nucleus reactions from subthreshold 
energies to a few GeV
employing empirical spectral functions~\cite{Sibirtsev1}. 
It was found that at
subthreshold energies the dominant contribution to 
$\rho$-meson production  stems from
secondary $\pi + N$ interactions while at energies above 2 GeV the primary
proton-nucleon channel becomes more important. In the latter study 
the cross section for $\rho$-meson
production from $\pi + N$ collisions was separated
into two parts: i) an exclusive cross section related
to binary processes ($\pi N \rightarrow \rho N$), which is dominant at 
energies close to the reaction threshold, and ii) the
inclusive production of $\rho$-mesons at higher energies which is
dominated by phase space and can be described by the LUND string model
\cite{Lund}.

The cross section for the exclusive reaction $\pi + N \to \rho + N$ 
can be obtained - in the limit of vanishing interference terms - 
via an incoherent summation over all baryonic
resonances decaying into $\rho N$ channels~\cite{Teis,CMK1,CMK2}
as
\begin{equation}
\label{resona}
{\sigma}_{\pi N \to \rho N}=\frac{\pi}{2 k^2}  
\sum_R (2I_R+1) \frac{{\Gamma}_{\pi N}{\Gamma}_{\rho N}}
{{\left(\sqrt{s}-M_R \right)}^2+{\Gamma}_R^2/4},
\end{equation}
where $M_R$ and ${\Gamma}_R$ are the mass and total
width of the baryonic resonance,  while
${\Gamma}_{\pi N}$ and ${\Gamma}_{\rho N}$  are the partial
width for the initial and final state, respectively. Here $k$ is
the pion momentum in the resonance frame of reference.
We account for  a momentum dependence of the resonance width 
as~\cite{Teis,Martin}
\begin{equation}
\label{width}
\Gamma (k) = {\Gamma}_0 {\left( \frac{k}{k_R} \right)}^{2l+1}
{\left( \frac{k_R^2+ {\delta}^2}{k^2+ {\delta}^2} \right)}^{l+1},
\end{equation}
where $k_R$ is the momentum at ${\sqrt{s}}=M_R$, $l$ is 
the minimal orbital angular momentum of the system  and
${\delta}^2 = 0.1$~GeV/c.

The properties of the baryonic resonances are 
available for resonances up to a mass of about 2~GeV~\cite{PartData}.
In order to describe the  exclusive $\pi + N \to \rho +N$ reaction 
at energies above $\sqrt{s} \simeq$ 2~GeV we incorporate additionally 
two 'effective' resonances with masses of 2.4 and 3.3~GeV; 
all resonance properties adopted are listed in Table.\ref{ta1}.

\begin{table*}[h]
\begin{center}
\caption{\label{ta1} Properties of baryonic resonances used in the
calculations. Apart from the known resonances we 
have added the 'effective' 
resonances $E1$ and $E2$ to describe properly the exclusive production
of $\rho$ mesons in $\pi N$ reactions.}
\vspace{0.6cm}
\begin{tabular}{|l|c|c|c|c|c|}
\hline
Resonance & $I_R$ & $M_R$ (MeV) & ${\Gamma}_R $ (MeV) & 
${\to}N{\pi}$ (\%) & ${\to}N{\rho}$ (\%) \\
\hline
$F_{15} $ & 5/2 & 1680 & 125 & 60 & 5 \\
$D_{13}$ & 3/2 & 1700 & 100 & 10 & 10 \\
$P_{11}$ & 1/2 & 1710 & 110 & 15 & 20 \\
$P_{13}$ & 3/2 & 1720 & 150 & 10 & 70 \\
$F_{17}$ & 7/2 & 1990 & 260 & 20 & 5 \\
$D_{33}$ & 3/2 & 1700 & 280 & 15 & 30 \\
$S_{31}$ &  1/2 & 1900 & 150 & 10 & 45 \\
$F_{35}$ & 5/2 & 1905 & 350 & 10 & 72 \\
$P_{31}$ & 1/2 & 1910 & 250 & 25 & 37 \\
$E1$ & 3/2 & 2400 & 1000 & 70 & 25 \\
$E2$ & 5/2 & 3300 & 1000 & 60 & 8 \\
\hline 
\end{tabular}
\end{center}
\end{table*}

\begin{figure}
\psfig{file=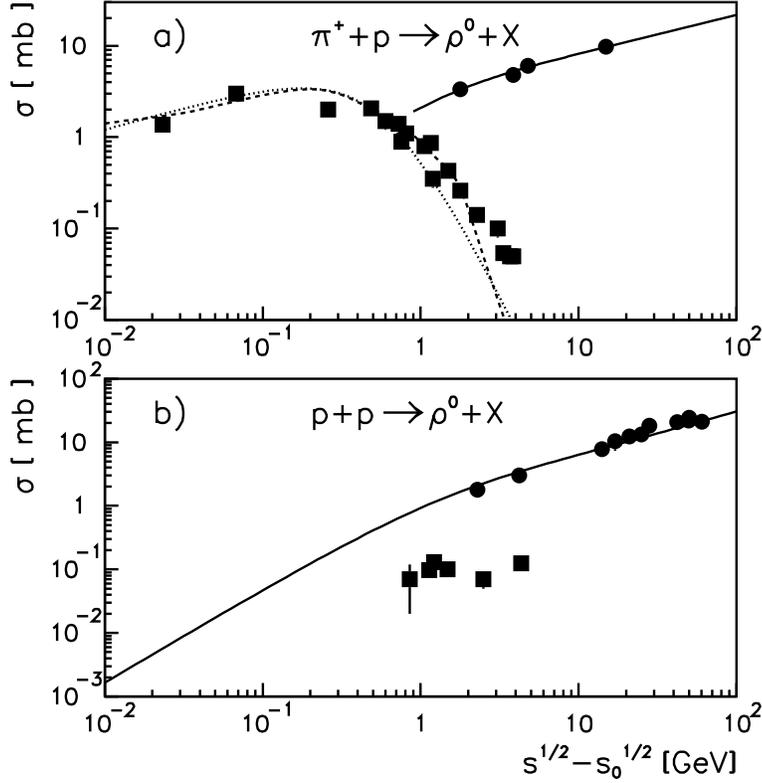,width=12cm}
\caption[]{\label{cbuu10} Cross sections for $\rho$-meson production
in $\pi^+ + p$ (a) and $p+p$ interactions (b). The solid lines show
the results from the LUND string model for the inclusive production 
cross section at higher energies, while the dashed line in (a) is
calculated within the effective resonance model~(\protect\ref{resona}).
The dotted line in (a) is the 
parametrization from~\protect\cite{Sibirtsev1}. 
The experimental data are taken from~\protect\cite{LA} 
and show the cross sections for the exclusive 
reaction $\pi^+ p \rightarrow \rho^+ p$ (full squares) (a) and 
$p p \rightarrow \rho^0 p p$ (full squares) (b) as well as  
for the inclusive reactions $\pi^+ p \rightarrow \rho^0 + X$ 
(full circles) (a)
and $p p \rightarrow \rho^0 + X$ (full circles) (b).}
\end{figure}

The experimental data for the exclusive reaction 
${\pi}^+ + p \to {\rho}^+ + p$~\cite{LA} are shown by full squares
in Fig.~\ref{cbuu10}a) as a function of 
$\sqrt{s}- \sqrt{s_0}$, with $\sqrt{s}$ denoting the invariant energy 
of the colliding particles while the threshold 
energy is given by $\sqrt{s_0}=m_N + m_{\rho}$.
The calculated cross section  within the resonance model 
(\ref{resona}) is shown in Fig.~\ref{cbuu10}a) by the dashed line. 
Note that for energies above $\sqrt{s}-\sqrt{s_0} \simeq$ 0.3~GeV a
reasonable reproduction of the experimental data is only possible
due to an incorporation of the two 'effective' 
resonances E1 and E2 as given
in Table 1. The dotted line in Fig.~\ref{cbuu10}a)  shows 
the parametrization from Ref.~\cite{Sibirtsev1} for comparison and
practically coincides with our present result from (\ref{resona}).

Since the available experimental data show no difference~\cite{Sibirtsev2}
between the reactions ${\pi}^- +p \to {\rho}^0 +n$,
${\pi}^- +p \to {\rho}^- +p$ and ${\pi}^+ +p \to {\rho}^+ +p$
we assume in the following that the
cross sections are the same for ${\pi}^+$, ${\pi}^-$ and ${\pi}^0$,  
for neutrons and protons as well as for ${\rho}^+$, ${\rho}^-$ and
${\rho}^0$-mesons, respectively.

At low energies above threshold only the exclusive 
reactions discussed above
are energetically allowed. However, an additional production of 
pions, i.e. $\pi + N \to \rho +N + \pi$'s,  is possible above
$\sqrt{s}-\sqrt{s_0} \ge m_\pi$.
Since there are not enough experimental data available for 
the inclusive $\rho$-meson production from 
$\pi + N$ collisions that allow to construct reliable parametrizations, 
we calculate the cross sections for the inclusive reactions
within the Lund-String-Model (LSM) from~\cite{Lund}.
The calculated results within the LSM can be fitted by 
\begin{equation}
\label{par1}
\sigma (\pi +p \rightarrow \rho+X) = 
3.6~[mb] \ \times  \ {\left( x-1 \right)}^{1.47} \ \times \ x^{-1.25} ,
\end{equation}
where $x = s / s_0$ and $s_0=(m_N+m_{\rho})^2$.
Our fit to the  LSM results is shown
by the solid line in Fig.~\ref{cbuu10}a) and
reasonably well reproduces the available inclusive data on 
$\rho$-meson production (full circles) at higher energies.

For the $\rho$-meson production from nucleon-nucleon
collisions we use the parametrization obtained within the LSM as
\cite{Sibirtsev1}
\begin{equation}
\label{par6}
\sigma (p +p \rightarrow \rho+X) = 
2.2~[mb] \ \times \ {\left( x-1 \right)}^{1.47} \ \times \  x^{-1.1} 
\end{equation}
with $x = s / s_0$ and $s_0=(2m_N+m_{\rho})^2$ for the nucleon-nucleon
channels.
The fit~(\ref{par6}) to the LSM results is shown by the solid line in 
Fig.~\ref{cbuu10}b) in comparison to the 
experimental data for the inclusive production of 
$\rho$-mesons (full circles) from~\cite{LA}. For comparison we
additionally show the cross section for the exclusive reaction 
$ p + p \to p + p + {\rho}^0$ (full triangles) from \cite{LA}, 
which about 1 
GeV above threshold is already small compared to the inclusive
cross section. Furthermore, we assume that the cross sections are isospin
independent for all possible nucleon-nucleon channels allowed by charge
conservation.

Fig.\ref{cbuu10} clearly demonstrates that the cross section
for $\rho$-meson production from pion induced reactions
is significantly larger than that from nucleon-nucleon
collisions at the same available energy above the threshold. 
As a consequence
the pion induced channels dominate for $\rho$ meson production also
in p + A reactions close to the absolute threshold energies as shown in
Ref. \cite{Sibirtsev1}. 

\subsection{$\rho N$ interactions}
The absorption of $\rho$-mesons in the nucleus due to the reactions
$\rho +N \to m\pi + N$ with $m \ge 1$ substantially attenuates 
the cross section and spectrum of $\rho$-mesons in p + A collisions.
To account for the $\rho$-meson interaction in finite nuclei we
need the $\rho +N$ cross section at nuclear densities from 0 to $\rho_0
\approx 0.17 fm^{-3}$. Since detailed Brueckner calculations for
these reactions in the medium are not available we assume that the
in-medium cross sections are approximately the same as 
those in the vacuum.
In the latter case the total and elastic $\rho N$ cross sections  can 
be calculated  within  the resonance model as
\begin{equation}
\label{res1}
{\sigma}_{ \rho N}=\frac{\pi}{6 k^2}  
\sum_R (2I_R+1) \frac{{\Gamma}_{\rho N}{\Gamma}_{out}}
{{\left(\sqrt{s}-M_R \right)}^2+{\Gamma}_R^2/4} ,
\end{equation}
where $M_R$, ${\Gamma}_R$ and ${\Gamma}_{\rho N} $ denote the mass, total
and partial width ($R \to \rho N$)
of the baryonic resonances, respectively. In Eq. (5)  
$k$ now is the momentum of the $\rho$-meson in the resonance frame.
The summation is performed again over all the baryonic resonances 
listed in Table.\ref{ta1} thus neglecting the interference
between the resonances. The $\rho +N$ total cross 
section is calculated with
${\Gamma}_{out}={\Gamma}_R$ while the elastic
cross section is evaluated with ${\Gamma}_{out}={\Gamma}_{\rho N}$.

\begin{figure}
\psfig{file=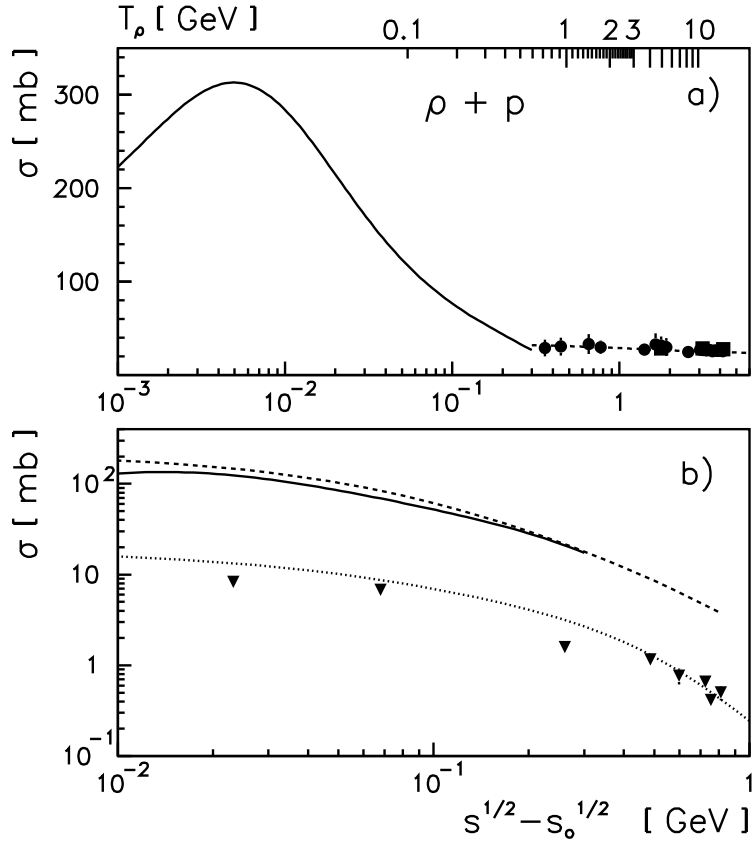,width=12cm}
\caption[]{\label{FI1} Cross sections for $\rho + p$ interactions:
a) shows the total cross section; 
the solid line is the calculation with the resonance 
model~(\protect\ref{res1}), while the dashed line indicates the
quark model relation~(\protect\ref{QCD1}). The solid circles show 
the experimental data extracted from 
vector dominance~(\protect\ref{vector}); 
the squares are the data from~\protect\cite{Anderson1}. 
b) The solid line shows
the elastic cross section from the resonance 
model~(\protect\ref{res1}), while the dashed line 
results from the $\sigma$-exchange calculations~(\protect\ref{exc1}).
The dotted line and the triangles indicate the cross section 
for the reaction
$\rho +N \to \pi +N$ obtained by detailed 
balance~(\protect\ref{detal},\protect\ref{detal1}).}
\end{figure}

The result of our calculations for the $\rho +N$ total cross section 
is shown in Fig.\ref{FI1}a) (solid line), while the elastic cross section 
is shown in Fig.\ref{FI1}b) (solid line);
both are displayed up to ${\sqrt{s}}-m_N-m_{\rho} \simeq 0.3$ GeV, only,
because we do not include in our calculation the heavier 
baryonic resonances
E1 and E2 due to possibly improper branching ratios. In view of the
neglect of interference terms and the limited energy range considered our
calculations should only be considered as an estimate that will have to be
improved in future. We note, however, that a momentum dependent width 
according to Eq.~(\ref{width}) compared to a fixed average width does not
change these cross sections in Fig. \ref{FI1} sizeably.

Following the quark model relations one can express the 
cross section for the  $\rho + N$ interaction at high energies 
in terms of pion-nucleon cross sections as
\begin{equation}
\label{QCD1}
{\sigma}_{ {\rho}^0 N}=\frac{1}{2} \left( {\sigma}_{ {\pi}^- N} +
{\sigma}_{ {\pi}^+ N} \right) .
\end{equation}
The total $\rho +N$ cross section according to
Eq.(\ref{QCD1}) is shown in Fig.\ref{FI1}a) by
the dashed line. For the $\pi N$ cross sections entering here we
have used a Regge fit to the experimental data from~\cite{PartData}.
Note, that there is a reasonable agreement between the results from
the resonance model and the quark model in the overlapping energy region.

The calculation of the $\rho +N \to \rho +N$ elastic scattering
in an effective meson-exchange model
may be restricted to a diagram with the $\sigma$-exchange \cite{Friman1}.
In the $t$-channel the elastic cross section then can be expressed as
\begin{equation}
\label{exc1}
{\sigma}_{\rho N \to \rho N}= \frac{1}{8\pi} \frac{m_N^2}{4s}
\int_{-1}^{1} dx \ \bar{M}(s,x) ,
\end{equation}
where $s$ is the square of the total center-of-mass energy,
$m_N$ is the nucleon mass and the isospin averaged squared
amplitude is given by~\cite{Friman1,Joos}
\begin{equation}
\label{exc2}
\bar{M}(s,x) = 4 \frac { g^2_{\sigma \rho \rho} g^2_{\sigma NN}}
{m_{\rho}^2}
\left( 1- \frac{q^2}{4m_N^2} \right)
{\left( \frac{1}{q^2-m_{\sigma}^2} \right)}^2
\left( m_{\rho}^4-\frac{m_{\rho}^2q^2}{3}+\frac{q^4}{24} \right)
F^2_{\sigma NN} F^2_{\sigma \rho \rho}
\end{equation}
where
\begin{equation}
q^2=\frac{\left[ s-(m_N+m_{\rho})^2 \right] \left[s-(m_N-m_{\rho})^2 
\right]}
{2s} (x-1) .
\end{equation}
In Eq.(\ref{exc2}) we use the monopole form factors
\begin{equation}
\label{exc3}
F_{\sigma NN}=\frac{{\Lambda}_1^2-m_{\sigma}^2}{{\Lambda}_1^2+q^2} ,
\ \ \ \ 
F_{\sigma \rho \rho}=\frac{{\Lambda}_2^2-m_{\sigma}^2}
{{\Lambda}_2^2+q^2} .
\end{equation}
The cut-off parameters ${\Lambda}_1=0.9$~GeV and
${\Lambda}_2=1.0$~GeV as well as the constant
$g^2_{\sigma \rho \rho}/4\pi =14.8$ were  taken from 
the recent analysis of
Friman and Soyeur~\cite{Friman2} on $\rho$-meson photoproduction
off nucleons, while  the coupling constant 
$g^2_{\sigma NN}/4\pi=8$ and $m_{\sigma}=550$~MeV were 
taken from the Bonn potential~\cite{Bonn}.
The elastic cross section calculated within the $\sigma$-exchange
model is shown in Fig.\ref{FI1}b by the dashed line and reasonably agrees
with that from the resonance model (solid line). Though the actual 
magnitude of this cross might change by about 30\% in a more sophisticated
analysis, it becomes clear that $\rho - N$ rescattering will be very 
important for the final $\rho$-meson spectra.

Furthermore, in a vector dominance model the photoproduction of a 
$\rho$-meson is related to the $\rho+N$ cross section as
\begin{equation}
\label{vector}
{\sigma}^2_{ \rho p} = \frac{{\gamma}^2_{\rho}}{4\pi}
\frac{64 \pi}{\alpha} 
\left. \frac{d\sigma_{\gamma p \to \rho p}} {dt} \right|_{t=0}
\end{equation}
where $\alpha$ is the fine-structure constant and the coupling
constant $ {\gamma}^2_{\rho}/4\pi =0.61$ 
according to Ref.\cite{Anderson}. 
The solid circles in Fig.\ref{FI1}a) show the total $\rho +p$ cross
section calculated from the experimental forward $\rho$-photoproduction 
amplitudes from~\cite{Anderson,Egloff,Ballam}. Furthermore, 
the squares in Fig.\ref{FI1}a) show the $\rho +N$ 
cross section extracted from $\rho$-photoproduction 
on the deuteron ($\gamma +d \to {\rho}^0 +d$)
independently of the vector dominance model~\cite{Anderson1}.

The cross section for the 
$\rho + N \to \pi + N$ reaction channel may be obtained exploiting 
detailed balance as
\begin{equation} 
\label{detal}
{\sigma}_{{\rho}^+ p \to {\pi}^+ p}= \frac {1}{3} \ \frac{q_{\pi}^2}
{q_{\rho}^2} \ \
{\sigma}_{{\pi}^+ p \to {\rho}^+ p}
\end{equation}
and is shown by triangles in Fig.\ref{FI1}b) using the experimental
cross section of the inverse reaction from Ref.~\cite{LA}.

Note that Eq.(\ref{detal}) is only valid in case of 
stable particles. For a
broad resonance as the $\rho$-meson it is easy to 
show~\cite{Danielewicz} that  
\begin{equation}
\label{detal1}
{\sigma}_{{\rho}^+ p \to {\pi}^+ p}= \frac {m_{\rho}}{6\pi} \ 
\frac{q_{\pi}^2}
{q_{\rho}} \ \
{\sigma}_{{\pi}^+ p \to {\rho}^+ p}
{\left( \int_{2m_{\pi}}^{\sqrt{s}-m_N} dW \ W \ q_f 
\frac{{\Gamma}_{\rho}}
{( W-m_{\rho})^2+{\Gamma}_{\rho}^2/4} \right)}^{-1} ,
\end{equation}
where $q_f$ is given as
\begin{equation}
q^2_f=\frac{\left[s-(m_N+W)^2 \right] \left[s-(m_N-W)^2 \right]}
{4s},
\end{equation}
and $m_{\rho}$, ${\Gamma}_{\rho}$ are the mass and width of the
$\rho$-meson, respectively. The cross section for 
the reaction ${\sigma}_{{\rho}^+ p \to {\pi}^+ p}$ calculated
by Eq.(\ref{detal1}) within the prescription~(\ref{resona}) 
for  ${\sigma}_{{\pi}^+ p \to {\rho}^+ p}$  is
shown in Fig.\ref{FI1}b by the dotted line and 
differs only slightly from the prescription~(\ref{detal}). 

Note, that the results obtained from quite different approaches are in
reasonable agreement with each other. 
At low energies $T_{\rho} \leq$ 200 MeV the contribution from the
inelastic reaction channels to the total cross section are
dominant and much larger than the elastic cross section.
Starting from a $\rho$ kinetic energy
of about 300 MeV the total cross section then 
stays constant at roughly 30~mb.
Following the quark model the inelastic reaction channels to the total 
cross section are dominant here, too.

\section{In-medium modifications of the $\rho$-meson}
In the literature there are several approaches on the in-medium 
modifications of the $\rho$-meson. Without going 
into a detailed discussion
we simply adopt two different models to explore the possibility if such
modifications can be observed in the two-pion decay channel. 

Following the results of Hatsuda~\cite{Hatsuda1}
obtained from QCD sum-rules the in-medium mass $m^{\ast}$ of the
$\rho$-meson can be approximated  as
\begin{equation}
\label{Hat}
m^{\ast} = m_B \ (1- \alpha {\rho}_N / {\rho}_0 )
\end{equation}
where $\alpha = [0.16 \pm 0.06]$, $m_B$=0.7699~GeV is the bare mass 
and ${\rho}_N $ is the
actual nuclear density, while ${\rho}_0=0.17$~fm$^{-3}$.

\begin{figure}
\psfig{file=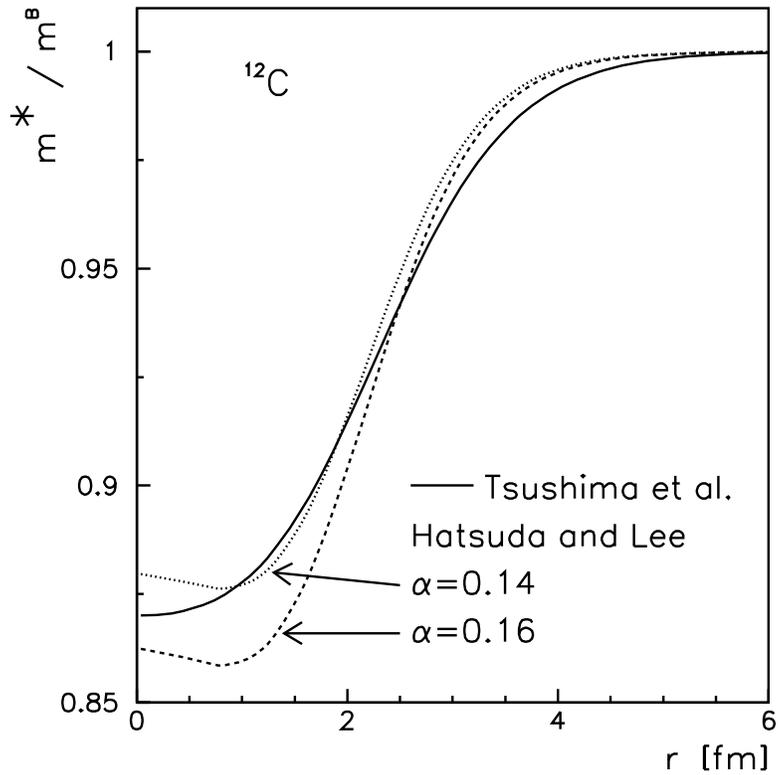,width=12cm}
\caption{\label{cbuu9} The  ratio of the
in-medium to the bare $\rho$-meson mass in a $^{12}C$-target
as a function of  the distance $r$ from the center of the nucleus.
The solid line shows the numerical 
results  from~\protect\cite{Tsushima1,Tsushima2}, the
dashed line is our result employing the
linear parametrization~(\protect\ref{Hat}) in a 
local density approximation
for  a parameter $\alpha$ = 0.16 in Eq. (\protect\ref{Hat}) 
while the dotted line is obtained for $\alpha$ = 0.14.}
\end{figure}

Recently, the $\rho$-meson mass in finite nuclei was calculated
by Saito, Tsushima and Thomas~\cite{Tsushima1} within the 
quark-meson coupling model. 
The ratio $m^{\ast} / m_B $ from~\cite{Tsushima1,Tsushima2} 
for $^{12}C$ is shown in 
Fig.\ref{cbuu9} by the solid line as a function of the radial distance r.
The dashed line in Fig.\ref{cbuu9} illustrates the density
dependence of the $\rho$-meson mass according to~(\ref{Hat}) for $\alpha$ 
= 0.16 using
the $^{12}C$ density distribution from~\cite{Tsushima1,Tsushima2}. 
The dotted 
line is obtained for $\alpha$ = 0.14, respectively. 
Note that the local density approximation works quite well 
in comparison to
the calculations from Ref. \cite{Tsushima1,Tsushima2}.
Indeed, the actual numerical results in \cite{Tsushima1,Tsushima2}
show only a weak deviation from the
linear parametrization~\cite{Hatsuda,Hatsuda2,Jin} for a fixed nuclear 
density. In view of the uncertainty in the parameter $\alpha$ it thus
appears justified to adopt the local density approximation furtheron.
Accordingly, in the following calculations we will use the 
scaling~(\ref{Hat}) while
keeping the width of the in-medium $\rho$-meson equal to the free one.

\begin{figure}
\psfig{file=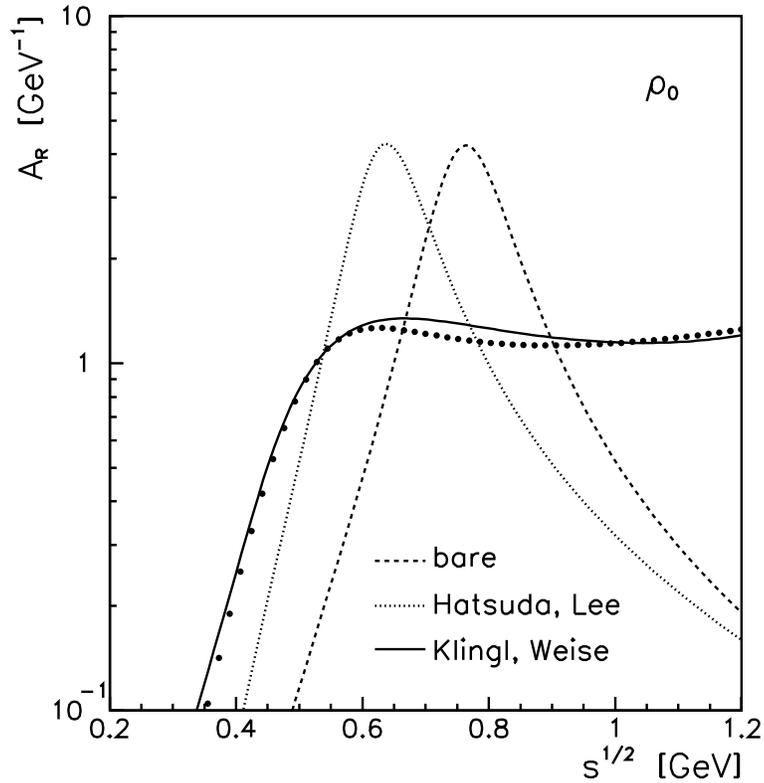,width=12cm}
\caption{\label{cbuu8} The spectral function of 
the $\rho$-meson from Klingl and 
Weise~\protect\cite{Klingl1,Klingl3} at normal 
nuclear matter density $\rho_0$
(solid line) in comparison with the
spectral function in free space (dashed line); the dotted line   
displays the $\rho$-meson mass distribution according to
Hatsuda and Lee~\protect\cite{Hatsuda} at $\rho_0$ for $\alpha$ = 0.16
employing the free width of the $\rho$ meson. The full dots show our
parametrization~(\protect\ref{wfit}) for the spectral 
function from Refs.~\protect\cite{Klingl1,Klingl3}.}
\end{figure}

Furthermore, according to the hadronic model calculations of Klingl and 
Weise~\cite{Klingl1,Klingl2} or Rapp, Chanfray and
Wambach~\cite{Wambach1,Wambach2} or Friman and Pirner~\cite{Friman5}
both the mass and the width of the $\rho$-meson are modified in
nuclear matter. In our present study we use the 
results from~\cite{Klingl1,Klingl2}.
Actually, for the $\rho \to \pi \pi$ decay mode 
we need the general spectral 
function $A_R = dM_R /dM_{{\pi}{\pi}}$,
i. e. the distribution of the resonance mass $M_R$  with respect to 
the invariant mass
of the two pions $M_{{\pi}{\pi}}$. The latter  is 
related to the quantity $R^{I=1}$ from~\cite{Klingl1} as
\begin{equation}
\label{weis}
A_R = {\left( \frac{g_{\rho}}{m_B^2} \right)}^{2} \ 
\frac{s^{3/2}}{6{\pi}^2} \ \times R^{I=1},
\end{equation}
where $g_{\rho}$ = 6.05 is the strong coupling constant, $m_B$ the
free $\rho$-meson mass and $\sqrt{s}$ the invariant mass of the two pions.
The solid line in Fig.\ref{cbuu8} shows the spectral 
function~(\ref{weis}) of the $\rho$-meson at 
density ${\rho}_N = {\rho}_0$ calculated 
with $R^{I=1}$ from Ref.~\cite{Klingl3}.

For numerical purposes we have parameterized~(\ref{weis}) 
at invariant energies above 2 $m_{\pi}$ in the form
\begin{equation}
\label{wfit}
A_R = \frac{{\Gamma}_{\rho}}{(\sqrt{s}-m^{\ast})^2+
{\Gamma}_{\rho}^2} \  a \ s^2
\end{equation}
with the parameters
\begin{eqnarray}
{\Gamma}_{\rho} &=& 2 {\Gamma}_0 \nonumber \\
m^{\ast} &=& m_B \ (1- 0.34 {\rho}_N / {\rho}_0 ) \\
a &=& 0.5 +0.5 {\rho}_N / {\rho}_0 , \nonumber
\end{eqnarray}
${\Gamma}_0=151$~MeV and $m_B$ being the bare mass
of the $\rho$-meson. The 
fit~(\ref{wfit}) is shown in Fig.\ref{cbuu8} by the full dots
and reasonably reproduces the spectral 
function from Ref.~\cite{Klingl1,Klingl3} at densities 
${\rho}_0/2 \le {\rho}_N \le {\rho}_0$.

The dashed line in Fig.\ref{cbuu8} represents the $\rho$-meson spectral 
function in free space as
\begin{equation}
\label{rhom}
A_R = \frac{1} {2 \pi}  \frac { {\Gamma}_{\rho}} 
{(\sqrt{s}-m_{\rho})^2 + {\Gamma}_{\rho}^2/4}
\end{equation}
with $m_{\rho}=m_B$  while ${\Gamma}_{\rho}$
is the momentum dependent $\rho$-meson width~(\ref{width}), 
respectively. Furthermore, 
the dotted line in Fig.\ref{cbuu8}  shows the spectral 
function~(\ref{rhom}) with the free $\rho$-meson width and 
an average in-medium mass $m_{\rho}=m^{\ast}$ from Eq.~(\ref{Hat}) 
for $\alpha$=0.16.

In the following calculations we will use the different spectral functions
in Fig.\ref{cbuu8} for a study of the $\rho$-meson 
production in $p+A$ collisions accounting for the in-medium
modifications as discussed above.

\section{Background from uncorrelated pions}
A proper calculation of the pion spectra from $p+A$
collisions is especially important for the study
of $\rho$-meson production via the two pion decay mode 
($\rho \to \pi \pi$), because multiple pion production from the reactions
$p+A \to m\pi + X$ with $m \ge 2$, which are not 
associated with $\rho$-meson
decays, substantially contribute to the invariant mass 
spectrum of two pions and have to be subtracted.

We use the coupled channel transport model \cite{CaB} for the description
of p + A reactions that allows to account for the
final state interactions of all hadrons and especially for those of 
the pions from the $\rho \to \pi \pi$ decay in the medium.
The transport approach \cite{CaB} is known to describe experimental data
on pion production in a wide range of bombarding 
energies~\cite{CaB,pions}. Here, we briefly present a comparison with the
experimental data available in the energy region of interest. Since the
semiclassical transport approach  \cite{CaB} involves a local density
approximation explicitly, this also serves as a test for the underlying
assumptions of the dynamical model itself. 

\begin{figure}
\psfig{file=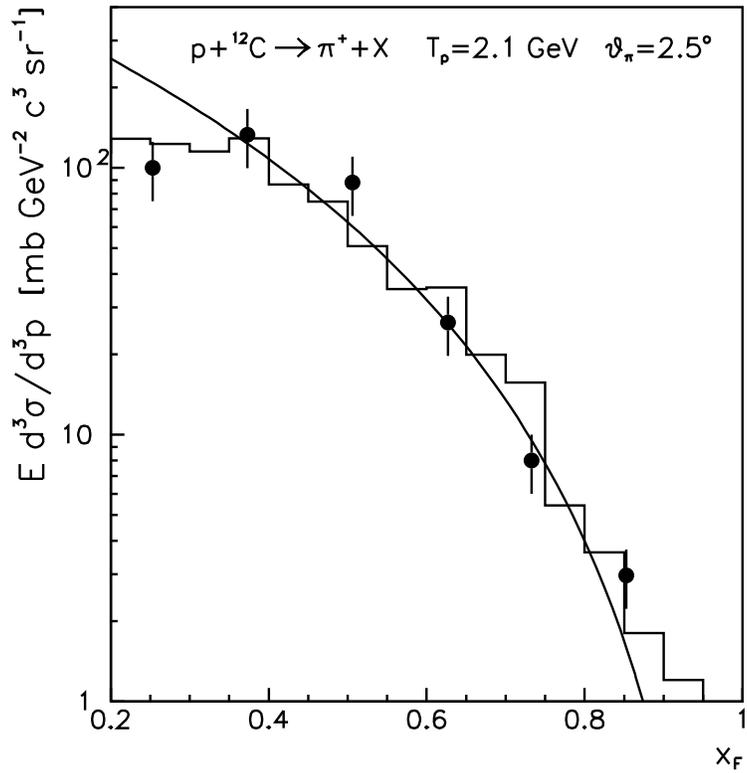,width=12cm}
\caption{\label{cbuu5} Invariant cross section for $\pi^+$-meson
production from $p+{^{12}C}$ collisions at 2.1 GeV at an emission
angle of 2.5$^o$ in the laboratory 
as a function of the Feynman variable $x_F$. The full dots show
the experimental data from~\protect\cite{Papp}; the
solid line is the scaling from~\protect\cite{Schmidt} while the histogram
shows our result calculated within the transport approach.}
\end{figure}

Fig.\ref{cbuu5} shows the spectrum of $\pi^+$-mesons produced
in $p+{^{12}C}$ collisions at  2.1~GeV  at the
pion emission angle of 2.5$^o$ in the laboratory. 
The data (full dots) from Ref. \cite{Papp} are
shown as a function of the Feynman variable 
$x_F=p_l^{\ast}/p_{max}^{\ast}$, where $p_l^{\ast}$ is
the longitudinal momentum of the pion in the center-of-mass
system of the incident proton and nucleon at rest while 
$p_{max}^{\ast}$ is defined as
\begin{equation}
p_{max}^{\ast} = 
\frac{{\left( [s-(m_N+m_{\pi})^2][s-(m_N-m_{\pi})^2] \right)}^{1/2}}
{2\sqrt{s}}.
\end{equation}  
The solid line shows
the scaling behaviour of the cross section from~\cite{Schmidt},
which describes $p+A$ reactions from a bombarding energy
of 1 GeV up to very high energies. The histogram represents our
calculations within the transport approach and 
shows a good agreement with 
the experimental results.

\begin{figure}
\psfig{file=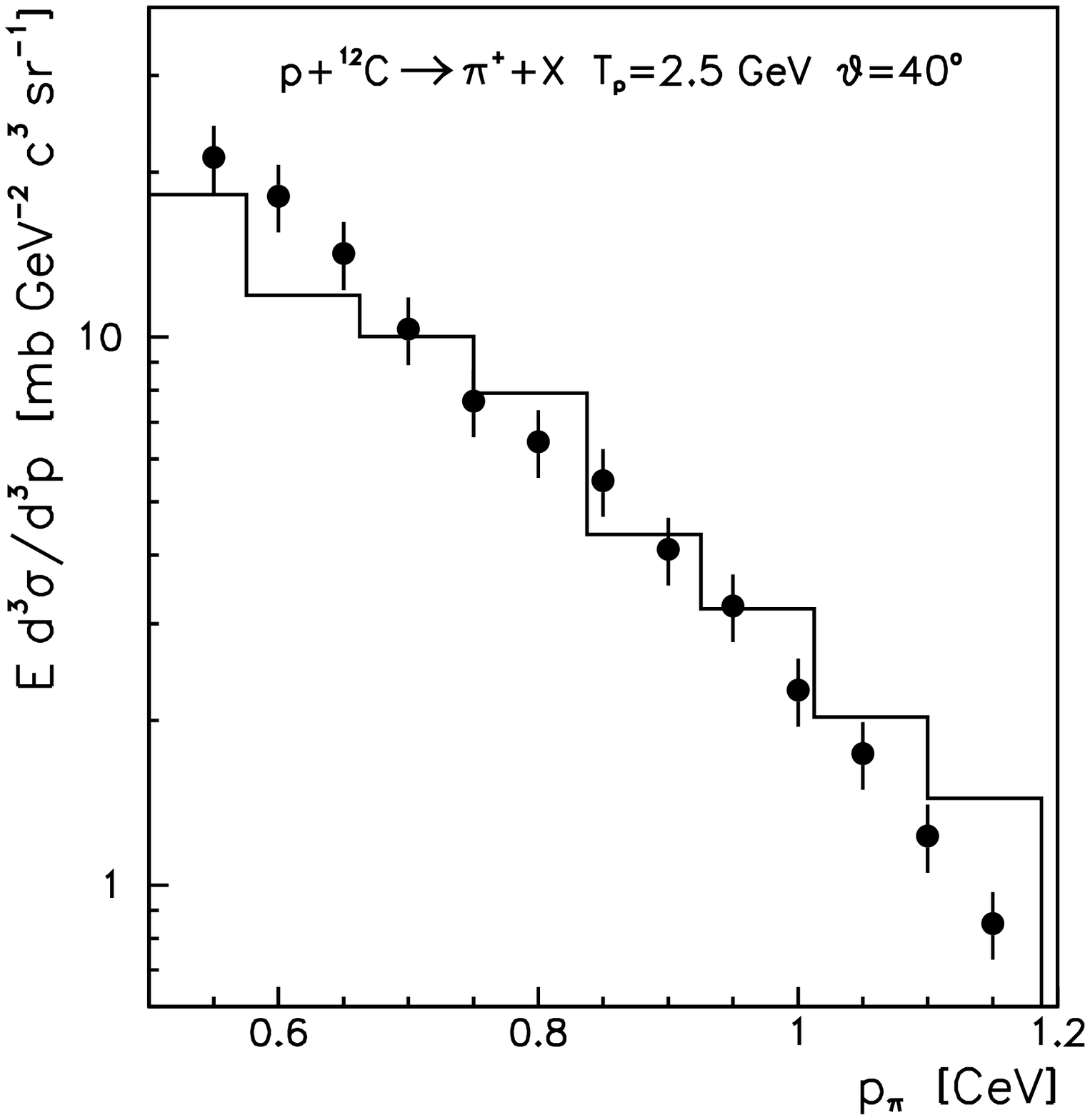,width=12cm}
\caption[]{\label{cbuu4} The invariant cross section for 
$\pi^+$-production from $p+{^{12}C}$ collisions at 
2.5 GeV at the pion emission angle of
40$^o$ in the laboratory. The experimental data~\protect\cite{Grosse1}
are shown by full dots while
the histogram represents the result from the transport calculations.}
\end{figure}

Fig.\ref{cbuu4} shows the invariant cross section for ${\pi}^+$-meson
production from $p+{^{12}C}$ collisions at 2.5~GeV
at an emission angle of 40$^o$ in the laboratory. 
The experimental data from the KaoS Collaboration~\cite{Grosse1}
are shown by the full dots while the transport calculations are
presented in terms of the histogram. We slightly overestimate the pion
spectrum at high momenta and underestimate for 
$p_{\pi} \approx$ 0.5 GeV/c.
These deviations are basically due to the local density approximation
employed in the transport model which is more uncertain in the surface of
a light nucleus such as $^{12}C$. Though keeping in mind these minor
deviations we conclude that 
the pion production and the proton-nucleus 
dynamics are described quite well
within the transport approach such that we now can proceed with 
the $\rho$-meson and two-pion dynamics.

\begin{figure}
\psfig{file=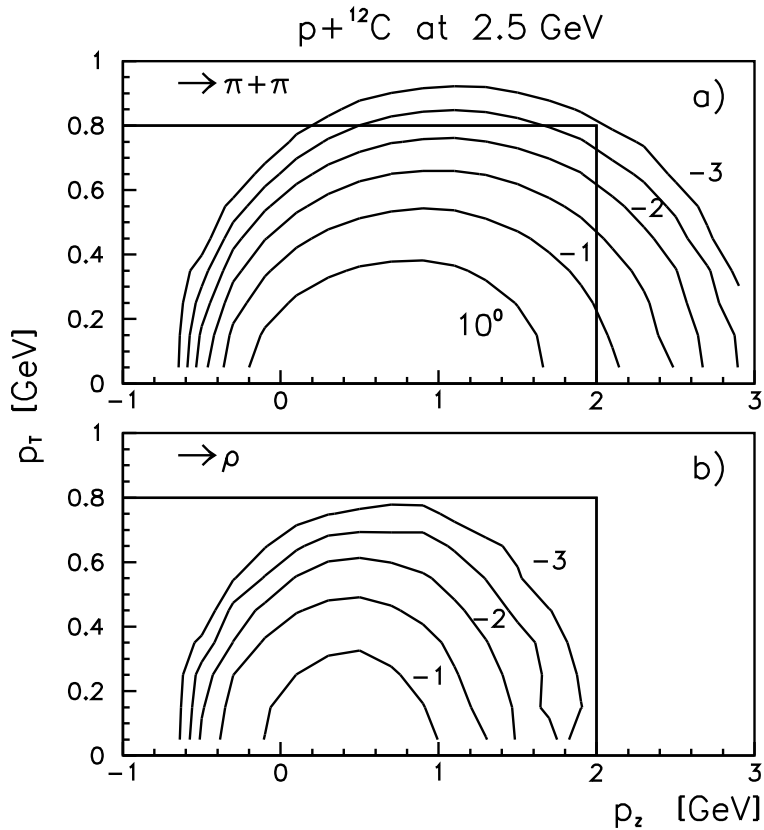,width=12cm}
\caption[]{\label{cbuu2} The 
longitudinal and transverse momentum distribution of pion pairs from 
$p+{^{12}C}$ collisions at 2.5~GeV; a) shows the results calculated for 
uncorrelated pion pairs while b) is for the pion pairs from $\rho$-decay.
The rectangles indicate the cuts used in our analysis.}
\end{figure}

Due to the different kinematics for the production of multipion states
and $\rho$-mesons from $p+A$ collisions the distribution of pion pairs
in momentum space should be much broader;
thus kinematical cuts could be used in order to enrich 
the contribution from 
$\rho$-meson decays. In this respect we show in 
Fig.\ref{cbuu2}a the distribution in 
longitudinal and transverse momentum of pion 
pairs from $p+{^{12}C}$ collisions
at 2.5~GeV, that do not stem from $\rho$-decays; 
Fig. \ref{cbuu2}b displays
the corresponding momentum distribution of pion pairs from $\rho$-decays,
respectively. As expected, the phase space of pion pairs from $\rho$-meson
decay is substantially smaller than that for the uncorrelated pions.
Using the kinematical cuts as shown in Fig.\ref{cbuu2}  
($p_z \leq $ 2 GeV/c, $p_T \leq$ 0.8 GeV/c) one can slightly 
enhance the $\rho$-meson signal in comparison to the background. In the 
following analysis we thus will always adopt the kinematical cuts 
$p_z \leq $ 2 GeV/c and  $p_T \leq$ 0.8 GeV/c.

\section{$\rho$-meson production from $p+A$ collisions.}
The inclusive production of $\rho$-mesons for $p+{^{12}C}$ 
reactions has been
calculated in Ref. \cite{Sibirtsev1} on the basis of empirical spectral
functions for energies from the absolute threshold up to a few GeV. Since
in this work we consider bombarding energies above 2 GeV the semiclassical
transport approach employed should yield similar inclusive  production
cross sections when employing the same cross sections for the $p N$ and
$\pi N$ production channels, respectively, and discarding any final state
interactions of the $\rho$-mesons. In fact, by comparing the inclusive
$\rho^0$ production cross sections in this limit from our 
transport approach 
with those from \cite{Sibirtsev1} for bombarding energies of 2 - 3 GeV
(cf. Table~\ref{ta2}) we find that
both models differ only by about 20 \%. The latter deviations between the
two approaches are due to the local density approximation and the on-shell
assumption in the transport model. Since these differences are lower than 
the uncertainty in the elementary production cross 
sections we can use the 
coupled channel transport approach quite confidently for 
the following analysis,
however, can include additionally all final state interactions of the 
$\rho$-mesons as well as for the pions which are more important in view of
their large cross sections with nucleons.

\begin{table*}[h]
\begin{center}
\caption{\label{ta2} Comparison of the inclusive ${\rho}^0$-production
cross section in mb for $p+{^{12}C}$ reactions within
the spectral function approach (SFA) \protect\cite{Sibirtsev1} and
our coupled channel transport approach (CBUU). For this comparison
all final state interactions of the $\rho$-mesons have been discarded.}
\vspace{0.6cm}
\begin{tabular}{|c|c|c|}
\hline
T (GeV) & SFA & CBUU \\
\hline
2.0 & 0.36 & 0.43 \\
2.5 & 1.2 & 1.5 \\
3.0 & 2.6 & 2.9 \\
\hline 
\end{tabular}
\end{center}
\end{table*}

\subsection{$\rho$-meson dynamics} 
\begin{figure}
\psfig{file=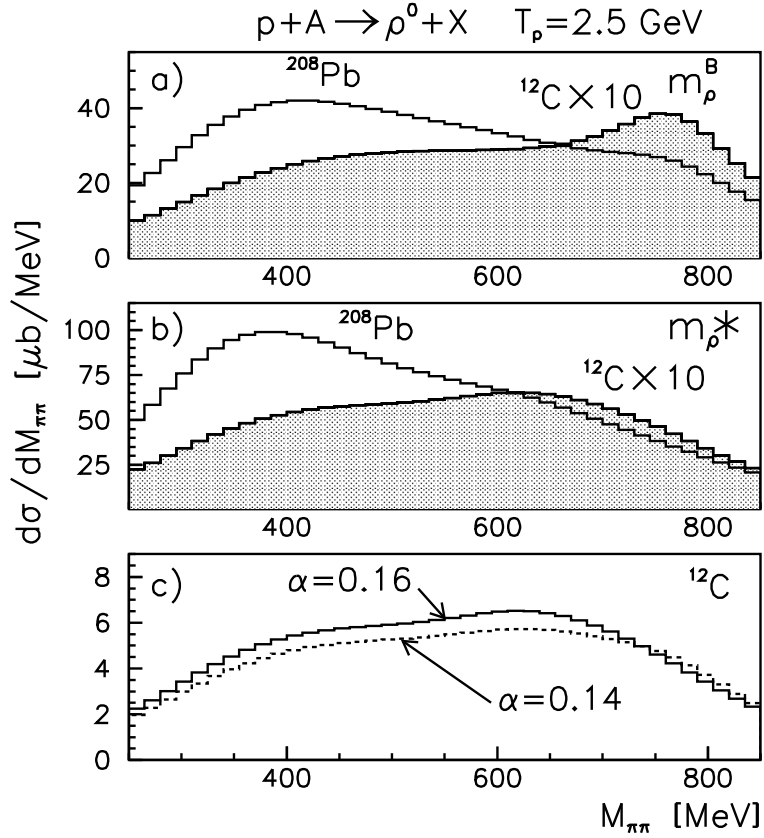,width=12cm}
\caption[]{\label{cbuu7}The invariant mass distribution 
of pion pairs from $\rho$-decays for
$p+{^{12}C}$ (hatched histogram) and $p+{^{208}Pb}$ collisions at 2.5 GeV 
calculated with the
bare (a) and in-medium (b) spectral function of the $\rho$-meson
according to Eq. (\protect\ref{Hat}) for $\alpha$ = 0.16; c) shows the
calculated result in the dropping mass scenario for $\alpha$ = 0.16 (solid
histogram) and $\alpha$ = 0.14 (dashed histogram), respectively.  
The spectra for $^{12}C$ are scaled by a factor of 10 in a) and b).}
\end{figure}

We first investigate the 'clean' $\rho^0$-meson 
production from $p+A$ collisions by neglecting 
the background from uncorrelated two
pion pairs. The spectra of $\pi^+\pi^- $ pairs are 
shown in Fig.\ref{cbuu7}
for $p+{^{12}C}$ and $p+{^{208}Pb}$ at a
bombarding energy of 2.5~GeV as a function of their invariant mass.
The calculations are performed for a bare $\rho$-meson spectral function
(Fig.\ref{cbuu7}a) and with an in-medium modification of the $\rho$-meson
(Fig.\ref{cbuu7}b) according to Eq.(\ref{Hat}) for $\alpha$ = 0.16. 
Note that the spectra for $p+{^{12}C}$ are displayed 
by hatched histograms and are
scaled by a factor of 10  in order to compare the results
for the light and the heavy target more directly. 

The low energy part of the invariant mass distribution reflects the
pion pairs from $\rho$-decays inside the nucleus that have suffered
strong  final state interactions. Obviously, such events are 
dominating for the heavy nucleus $^{208}Pb$. 
Even without medium modifications
the $\rho$-resonance is practically melted and the mass distribution is 
close to a thermalized spectrum. For $^{12}C$ 
the rescattering of pions from $\rho$-decays
is significantly reduced and a 'bare' $\rho$-meson signal or an in-medium
$\rho$-meson signal can still be identified. In Fig.\ref{cbuu7}c) we
additionally show the calculated results for the invariant 
mass distribution
of the pions from $\rho$ decay for a $^{12}C$ target 
employing the parameters
$\alpha$ = 0.16 and 0.14, respectively, as in Fig. 3. 
The direct comparison
demonstrates the relative sensitivity to the parameter $\alpha$ in the
dropping mass scheme of Eq. (15); furthermore, it also 
demonstrates the relative
uncertainty arizing from the local density approximation in the transport
approach since in view of Fig. 3 this uncertainty can be simulated by a
corresponding change of the $\alpha$ parameter.

This initial study shows that the pion mode has substantial 
disadvantages compared to the dileptonic $\rho$-meson
decay not only for the investigation of the in-medium properties
of the $\rho$-meson but also for a conventional analysis of
$p+A \to \rho +X$ reactions that rely on the reconstruction of the
$\rho$-meson  via two pion invariant mass distributions.
The lightest target nuclei, however, may be used if
the invariant mass spectrum is measured with a sufficiently high 
accuracy.

\begin{figure}
\psfig{file=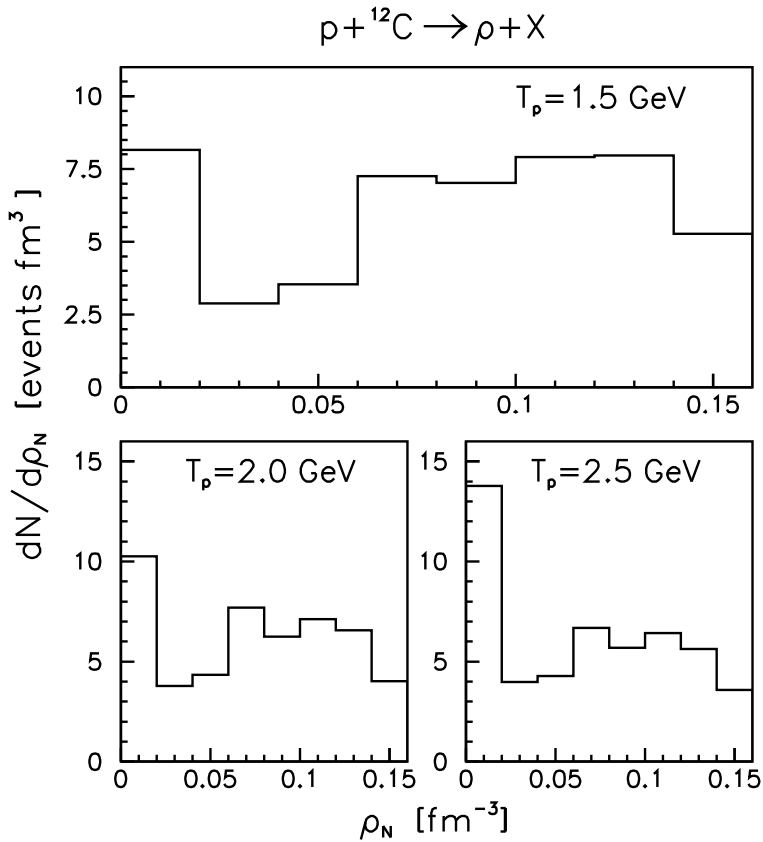,width=12cm}
\caption[]{\label{cbuu13} The distribution of the
$\rho$-decay rate as a function of the nuclear density
for $p+{^{12}C}$ collisions at bombarding energies 
of 1.5, 2.0 and 2.5~GeV.}
\end{figure}

Of fundamental interest for the $\rho$-meson dynamics - relevant
for in-medium modifications - is the distribution of the
decay rate as a function of the nuclear density. Obviously,
this distribution depends on the beam energy as well as on the
size of the target and should be properly calculated within the
transport approach. In this respect we display in Fig.\ref{cbuu13} the 
number of $\rho$-mesons decaying at the nuclear 
density ${\rho}_N$ for $p+{^{12}C}$ collisions at bombarding
energies of 1.5, 2.0 and 2.5 GeV. All distributions are
normalized to 1 in order to illustrate the relative variation with the
beam energy.

Note that events with densities ${\rho}_N \leq$~0.05~fm$^{-3}$
can be associated with a $\rho$-meson decay at the periphery of the target
or in the vacuum and carry no information about the medium 
effects. With increasing bombarding energy the nucleus
becomes more transparent for the propagation of $\rho$-mesons due to
lower total $\rho N$ cross sections and due to the fact that 
more $\rho$-mesons decay in the vacuum. 
Thus low energy collisions are more suitable to study the medium 
modification of $\rho$-meson properties, however, one should
keep in mind that the cross section for $\rho$-meson production at low 
energies is substantially smaller.

\begin{figure}
\psfig{file=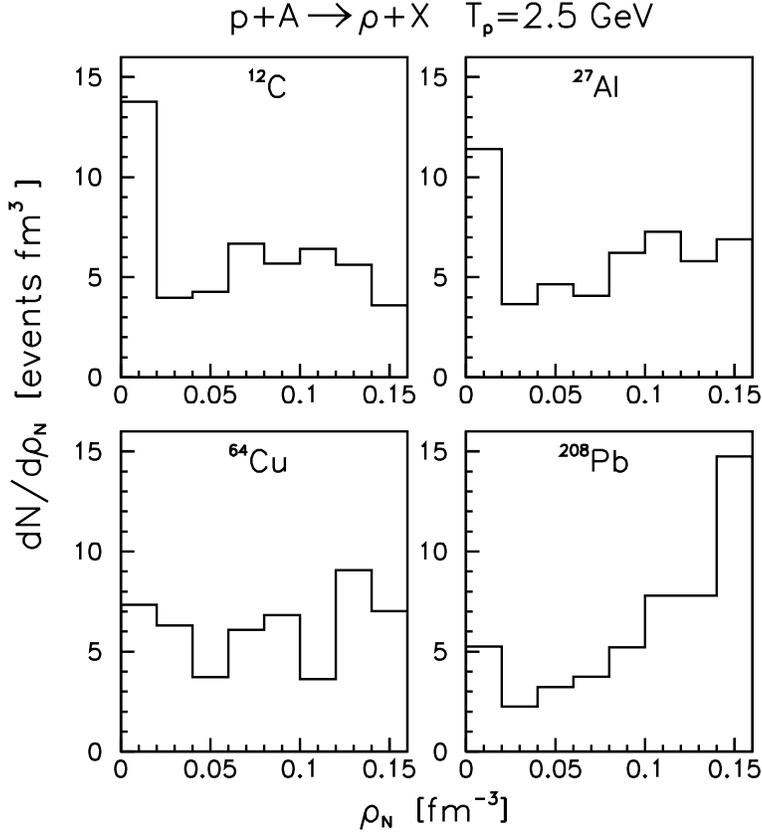,width=12cm}
\caption[]{\label{cbuu14} The $\rho$-meson
decay rate as a function of the nuclear density
for different targets at a bombarding energy of 2.5~GeV.}
\end{figure}

Fig.\ref{cbuu14}, furthermore,  shows the $\rho$-decay rate as a function
of the nucleon density at 2.5 GeV for the targets $^{12}C$, 
$^{27}Al$, $^{64}Cu$
and $^{208}Pb$. While for the $Pb$-target a major 
fraction of $\rho$-mesons
decays at normal nuclear matter density, this fraction decreases with the
target mass significantly. However, even for $^{12}C$ still 50\% of the
$\rho$'s decay at densities $0.5 \rho_0$-$\rho_0$ such that modifications
of the $\rho$-spectral function in the medium can still be 
tested with $^{12}C$.

\subsection{Calculations with a bare $\rho$ spectral function}
\begin{figure}
\psfig{file=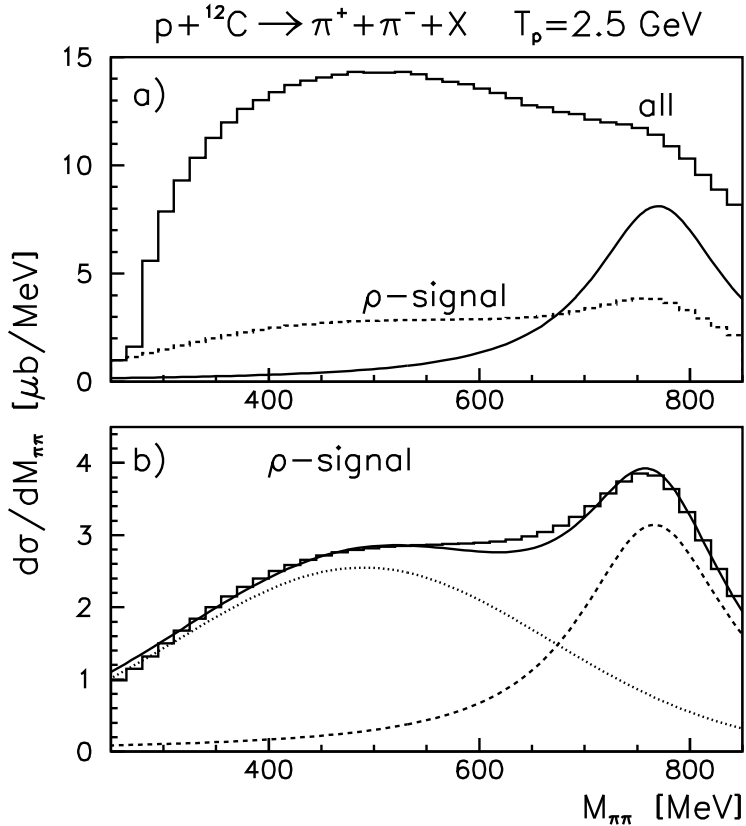,width=12cm}
\caption[]{\label{cbuu3}The invariant mass distribution of two pions from
$p+{^{12}C}$  collisions at 2.5 GeV using the bare spectral function
of the $\rho$-meson for $p_z \leq$ 2 GeV/c and $p_T \leq$ 0.8 GeV/c.
In a) the solid histogram shows the total 
$\pi -\pi$ invariant mass spectrum while the dashed 
histogram indicates the 
$\rho$ signal after subtraction of the combinatorial
background. The solid line is a Breit-Wigner distribution with the bare
$\rho$-meson properties normalized to the total $\rho$-production
cross section from the transport approach as expected in case of no final
state interactions.  In  b) the  histogram
shows the $\rho$-meson signal from a) while the dotted line is 
the statistical spectrum;
the dashed line is a  Breit-Wigner function with the $\rho$-meson
properties $M_0=766.9$~MeV and ${\Gamma}_0=173$~MeV and 
the solid line shows
the sum as discussed in the text.}
\end{figure}

In actual experiments the invariant mass distribution of pion
pairs from $\rho$-decay as presented in Fig. 8 is swamped by 
the background from pion pairs that stem from the reactions 
$p N \rightarrow \Delta \Delta \rightarrow
2 \pi N N$ and $p N \rightarrow R N \rightarrow 2 \pi N N$, where $R$ is
an intermediate resonance that decays into the 
two pion channel (cf. Fig.7a). 
The actual invariant mass distribution of pion pairs to be expected 
experimentally  
(within the cuts $p_z \leq$ 2 GeV/c and $p_T \leq$ 0.8 GeV/c) 
is shown in Fig.\ref{cbuu3}a) for $p+{^{12}C}$
at 2.5 GeV when neglecting any medium modifications of the $\rho$-meson.
The solid histogram in Fig.\ref{cbuu3}a represents the pion 
pair mass distribution 
including the  background pions as well as those from $\rho$-decay while
the solid curve indicates the contribution from $\rho$-decay 
as obtained from the total $\rho$ production cross section 
without any final state
interactions. The latter solid line thus would represent a maximum $\rho$
signal if all $\rho$-mesons would propagate out of the 
nucleus and decay in
vacuum which, of cause, is unrealistic.

In order to subtract the background  from uncorrelated pion pairs
we adopt a combinatorial method, i.e. mix pions from different 
events. This procedure was demonstrated in a recent experimental study of 
${\Delta}(1232)$-resonance excitation in heavy-ion collisions
by Badala et al.~\cite{Badala} 
to be very effective for the extraction of a true
correlation signal. The dashed histogram in Fig.\ref{cbuu3}a)
shows the difference between the real and mixed invariant mass
spectra and corresponds to pion pairs from $\rho$-decays including all
final state interactions of the $\rho$-meson as well as its decay pions
(cf. Fig. 8a). As discussed above the deviation of the 
solid line in Fig. 11a) from the dashed histogram is only 
due to $\rho$ and pion final state 
interactions. Consequently, the $\rho$ spectral function 
appears substantially
distorted by conventional hadronic interactions. 

According to Figs.\ref{cbuu13},\ref{cbuu14} a sizeable fraction 
of the $\rho$-mesons decays at very
low nuclear density for $^{12}C$ or in the vacuum where 
its spectral function
is well known. We thus can describe this 'outside' 
component approximately
by a Breit-Wigner function
\begin{equation}
\label{bw}
\frac{d\sigma}{dM_{\pi \pi}} =
\frac {C_0} {(M_{\pi \pi}-M_0)^2 + {\Gamma_0}^2/4} ,
\end{equation}
where $C_0$ is some constant which (in absence of 
$\rho$-meson melting)  is related to the total cross section of 
$\rho$-production from $p+{^{12}C}$ collisions 
while $M_0$ and ${\Gamma_0}$
are the mass and width of the bare $\rho$-meson, respectively.

The $\pi$-$\pi$ mass distribution, that is associated to the
$\rho$-meson production from $p+{^{12}C}$ collisions
(after background subtraction), is shown again in 
Fig.\ref{cbuu3}b) by the solid histogram. Apart from the 'outside'
decay contribution a further fraction of $\rho$'s 
decays 'inside' the medium
at nonzero density. Tentatively, we analyze this 'inside' component 
(together with all effects from final state interactions) 
in terms of a 'thermodynamical approach'. In this limit one can describe 
this part within the statistical model similar to the formalism 
developed in~\cite{Gale,Koch} by
\begin{equation}
\label{stat}
\frac{d\sigma}{dM_{\pi \pi}} =
N \ exp \left( - \frac{[M_{\pi \pi}-E_0]^2}{T_0^2} \right),
\end{equation}
where the normalization $N$, average excitation energy $E_0$ and
dispersion $T_0$ can be fitted to the $\rho$-mass spectrum for 
lower invariant mass $M_{\pi \pi}$
as shown in Fig.\ref{cbuu3}b) by the dotted line.
Accordingly, the solid line in Fig.\ref{cbuu3}b) 
shows our fit to the $\rho$-meson
signal by the sum of the  Breit-Wigner function and statistical 
spectrum~(\ref{stat}), 
where the first one involves the 
parameters $M_0$=766.9~MeV and ${\Gamma_0}$=173~MeV and is shown
by the dashed line in Fig.\ref{cbuu3}b). 

So far, we have still discussed the results calculated with 
the bare $\rho$-meson properties. The final state interactions
of $\rho$-mesons and pions from the $\rho \to \pi \pi$ decay yield
an invariant mass distribution which is close to the
recent predictions from Klingl and Weise~\cite{Klingl1} as discussed in 
Section~4 (cf. also Fig. 4). Any additional in-medium effects thus should
show up in a further distortion of the invariant mass distribution.

\subsection{In-medium modifications}
\begin{figure}
\psfig{file=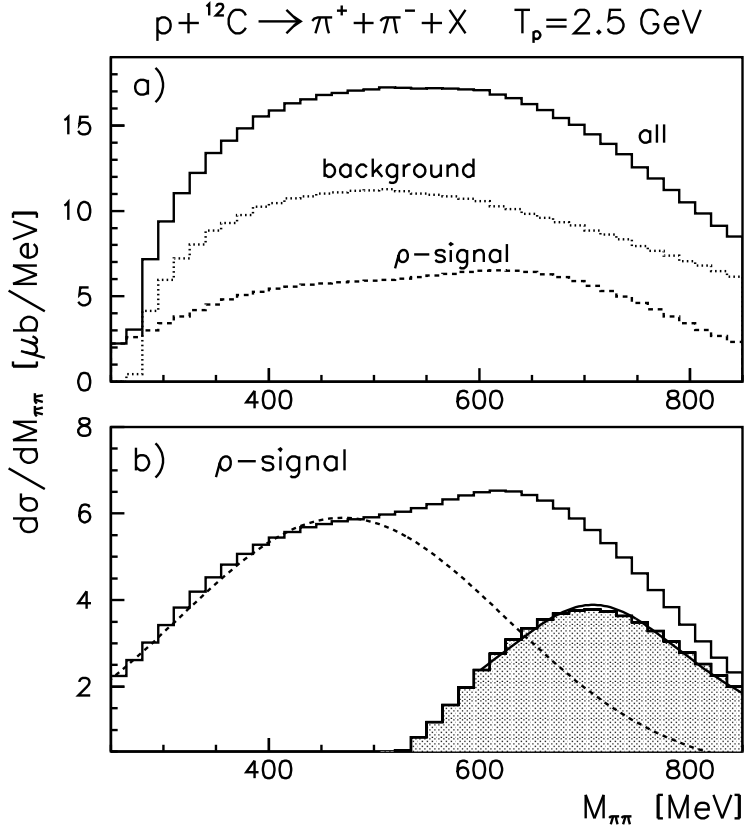,width=12cm}
\caption[]{\label{cbuu6}The invariant mass distribution of two pions from
$p+{^{12}C}$  at 2.5 GeV employing the in-medium
modification of the  $\rho$-meson according to Hatsuda and Lee 
\protect\cite{Hatsuda} for $\alpha$ = 0.16.
In a) the solid histogram shows the total 
$\pi -\pi$ invariant mass spectrum; the dotted histogram displays 
the background from uncorrelated pion pairs while
the dashed histogram gives the 
$\rho$-mass distribution after subtraction of the 
background.  In b) the solid histogram
shows again the $\rho$-meson signal from a); 
the dashed line is a fit with Eq.(\protect\ref{stat}) while  
the hatched histogram is the difference between the $\rho$ signal and 
the fit Eq.(\protect\ref{stat}), which can be described again by a
Breit-Wigner function (solid line).}
\end{figure}

As discussed above the QCD sum-rule approach 
by Hatsuda and Lee \cite{Hatsuda}
predicts a shift of the $\rho$-mass with density 
according to Eq. (\ref{Hat})
which corresponds to a shift of the $\rho$-meson pole. 
In order to explore
if such an effect could be seen in the two pion decay mode 
we have performed calculations for 
$p+{^{12}C}$ at 2.5 GeV employing Eq. (\ref{Hat}).
The solid histogram in Fig.\ref{cbuu6}a) shows the total
distribution in the invariant mass of two pions while 
the dotted histogram 
indicates the background obtained by mixing of pions from
different events. The latter is practically the same as that for those
calculations performed with the bare $\rho$-meson properties 
(cf. section 5.2).
The dashed histogram in Fig.\ref{cbuu6}a) represents the $\rho$-signal as
obtained by background subtraction.
It is seen that not only the shape of the mass
spectra, but also the absolute normalization are 
different from the calculations with the bare $\rho$-properties.
The $\rho$-meson signal is shown again in Fig.\ref{cbuu6}b) in
terms of the solid histogram, which at first 
glance is  quite complicated to analyze in order to extract
in-medium properties of the $\rho$-meson.

We proceed as in section 5.2 
and  fit the spectrum with the function~(\ref{stat}) 
as shown in Fig.\ref{cbuu6}b) by the dashed line.
The hatched histogram in Fig.\ref{cbuu6}b) then shows 
the difference between 
the $\rho$-mass spectrum and distribution~(\ref{stat}); 
it can again be fitted
by a Breit-Wigner distribution~(\ref{bw}) with $M=$708~MeV and 
${\Gamma}$=270~MeV. In comparison with Fig.\ref{cbuu3}b)
the dropping of the  $\rho$-meson
mass thus appears possible to be extracted from the invariant mass
distribution of two pions at least for light nuclei as $^{12}C$.

\begin{figure}
\psfig{file=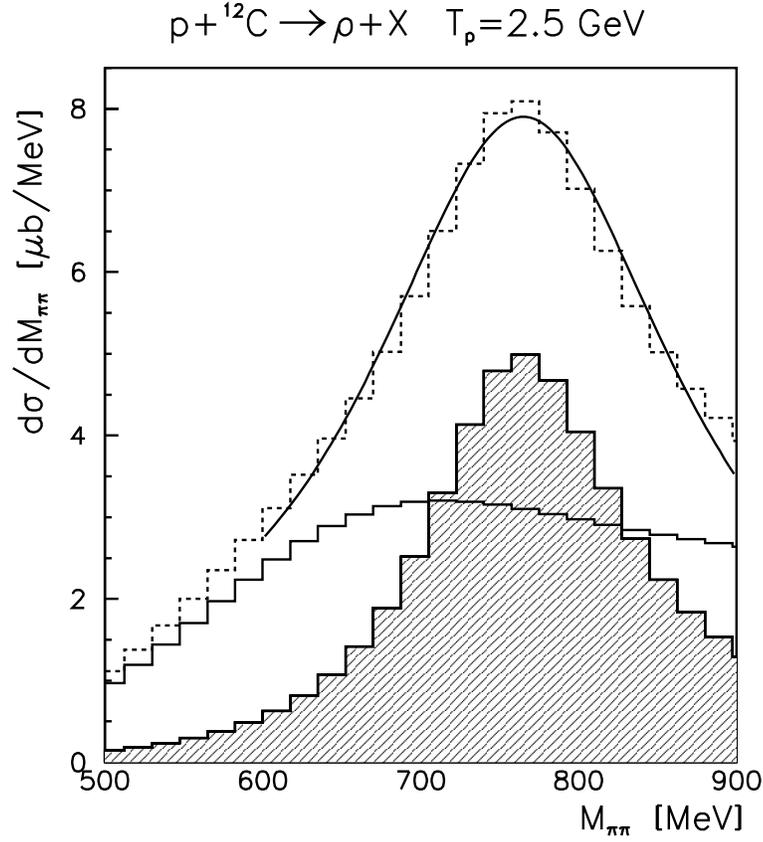,width=12cm}
\caption[]{\label{cbuu15}The invariant mass 
distribution of pion pairs from
$p+{^{12}C}$ at 2.5 GeV including the in-medium
modification of the  $\rho$-meson according to Klingl and 
Weise~\protect\cite{Klingl1,Klingl3}.
The hatched histogram shows the spectrum from $\rho$-mesons
decaying 'outside' the nucleus, the
solid histogram is the 'inside' component, while the dashed histogram 
is the sum of both. The solid line shows a Breit-Wigner 
fit with $M$= 765 MeV
and $\Gamma$=240 MeV.}
\end{figure}

We have also studied the modification of the $\rho$ spectral function
following the model from Klingl and Weise~\cite{Klingl1,Klingl3}.
For simplicity of the analysis we have 'switched-off' the final state
interactions of pions stemming from $\rho$-decay in order to
distinguish the change of the spectral function 
due to such processes from the in-medium 
modification~(15). Fig.\ref{cbuu15} shows our results 
calculated for $p+{^{12}C}$ collisions at 2.5~GeV. The hatched histogram
stems from $\rho$-mesons, which decay outside  or
at the periphery of the nuclear target, while the 
solid histogram illustrates
the pion pairs from the 'inside' decay of the $\rho$-mesons. Note
that now the internal component has no pronounced 
resonance structure anymore and is similar to the spectrum from pion 
rescattering. 
The dashed histogram in Fig.\ref{cbuu15} shows the sum of these two
components and the solid line is a fit~(\ref{bw}) with the parameters
$M=$765~MeV and ${\Gamma}$=240~MeV. The 'extracted' mass of the 
$\rho$-meson in this case is close to the bare mass while the width is 
considerably larger.

\section{Summary}
Within the coupled channel transport approach we have studied 
the production of $\rho$-mesons in proton-nucleus collisions and 
the detection of $\rho$-mesons via the pion decay mode 
($\rho \to \pi +\pi$)
employing different spectral functions for the in-medium $\rho$-mesons.

The elementary cross sections for $\rho$-production from 
$p+N$ and $\pi +N$
interactions as well as total and elastic $\rho +N$
final state interactions were calculated within the resonance model at low
invariant energies and in quark-string models at 
higher invariant energies, 
which provided consistent results in a reasonable agreement with
available experimental data. Note, however, that we 
have discarded interference
effects in the resonance model and that more elaborate approaches will be
necessary in future to pin down these cross sections more accurately.

Special attention was paid to the background from the
production of two uncorrelated pions as well as for the
attenuation of the $\rho$-meson yield due to $\rho N$ 
interactions as well as
the final state interactions ($\pi + N$) of pions emerging from
$\rho$-decay. We found that only light nuclear targets might be suitable
for the experimental study of $\rho$-production via the
pion decay  mode. Appropriate kinematical cuts and background subtractions
in principle allow to reconstruct the mass and width of the $\rho$-meson.
However, for such a study a good mass resolution for the pion 4-momenta is
required as well as a large detector acceptance. 

We have investigated different schemes for the modification of
the $\rho$-meson in nuclear matter and its consequences for
the invariant mass distribution of two pions that can be detected
asymptotically. The $\pi - \pi$ invariant mass 
spectra calculated within the hadronic
scenario of Klingl and Weise~\cite{Klingl1,Klingl2} are substantially
different from those obtained by simulations based 
on the QCD sum-rule approach  by Hatsuda and 
Lee~\cite{Hatsuda}. These differences are much larger 
than the theoretical 
uncertainties arizing from the local density approximation in the
transport approach and might actually 
be studied experimentally by large area pion spectrometers e.g. at
COSY, CELSIUS or the SIS. However, according to the authors opinion the
actual shape of the $\rho$-meson spectral function cannot be determined
accurately enough from the data analysis.

\vspace{1cm}
The authors gratefully acknowledge stimulating discussions with
B.~Friman, C.M.~Ko, S.~Leupold, U.~Mosel and  K.~Tsushima.
Furthermore, they like to thank F.~Klingl, W.~Weise, K.~Saito, K.~Tsushima 
and A.W.~Thomas for providing their numerical results.

\end{document}